\def\ONECOLUMN{}
\ifdefined\ONECOLUMN
	\documentclass[journal, 12pt, onecolumn, draftclsnofoot, a4paper]{IEEEtran}
\else
	\documentclass[journal, a4paper]{IEEEtran}
\fi
\IEEEoverridecommandlockouts
\usepackage{fancyhdr}
\usepackage{cite}
\usepackage{graphicx}
\usepackage{psfrag}
\usepackage{subfigure}
\usepackage{url}
\usepackage{balance}
\usepackage{stfloats}
\usepackage{amsmath,amsthm,amssymb}

\DeclareMathOperator*{\argmin}{arg\,min}
\usepackage{array}
\usepackage{fancyhdr}
\usepackage{float}
\usepackage{epsfig}
\usepackage{color}
\usepackage[nomain,acronym,toc]{glossaries}
\usepackage{float}

\newtheorem{remark}{Remark}

\theoremstyle{definition}

\newacronym{musimo}{MU-SIMO}{multiuser single-input multiple-output}
\newacronym{ap}{AP}{access point}
\newacronym{ils}{ILS}{integer least-squares}
\newacronym{snr}{SNR}{signal-to-noise ratio}
\newacronym{wl}{WL}{widely linear}
\newacronym{mlsd}{MLSD}{maximum-likelihood sequence detection}
\newacronym{map}{MAP}{maximum a posteriori}
\newacronym{mmse}{MMSE}{minimum mean-square-error}
\newacronym{ut}{UT}{user terminal}
\newacronym{bs}{BS}{base station}
\newacronym{ann}{ANN}{artificial neural networks}
\newacronym{dnn}{DNN}{deep neural network}
\newacronym{snn}{SNN}{super neural networks}
\newacronym{gmnn}{G-MNN}{giant modular neural network}
\newacronym{mnn}{MNN}{modular neural network}
\newacronym{pic}{PIC}{parallel interference cancellation}
\newacronym{mimo}{MIMO}{multiple-input multiple-output}
\newacronym{mud}{MUD}{multiuser detection}
\newacronym{amp}{AMP}{approximate message passing}
\newacronym{zf}{ZF}{zero forcing}
\newacronym{mf}{MF}{matched filter}
\newacronym{cdma}{CDMA}{code division multiple access}
\newacronym{noma}{NOMA}{non-orthogonal multiple access}
\newacronym{sic}{SIC}{successive interference cancellation}
\newacronym{cdf}{CDF}{cumulative distribution function}
\newacronym{ncl}{NC-Learning}{non-coherent learning}
\newacronym{cel}{CE-Learning}{channel equalized learning}
\newacronym{dcl}{DC-Learning}{direct coherent learning}
\newacronym{bp}{BP}{back-propagation}
\newacronym{vq}{VQ}{vector quantization}
\newacronym{awgn}{AWGN}{additive white Gaussian noise}
\newacronym{ai}{AI}{artificial intelligence}
\newacronym{dsp}{DSP}{digital signal processing}
\newacronym{csi}{CSI}{channel state information}
\newacronym{nn}{NN}{nearest-neighbor}
\newacronym{clnn}{CL-NN}{cluster-level nearest neighbor}
\newacronym{clknn}{CL-kNN}{cluster-level kNN}
\newacronym{ber}{BER}{bit error rate}
\newacronym{mai}{MAI}{multiple access interference}
\newacronym{lte}{LTE}{Long Term Evolution}
\newacronym{nr}{NR}{new radio}
\newacronym{sd}{SD}{sphere decoding}
\newacronym{lr}{LR}{lattice-reduction}
\newacronym{urllc}{URLLC}{ultra-reliable low-latency communication}
\newacronym{csit}{CSIT}{channel state information at transmitter}
\newacronym{csir}{CSIR}{channel state information at transmitter}
\newacronym{gpu}{GPU}{graphics processing unit}
\newacronym{pep}{PEP}{pairwise error probability}
\newacronym{relu}{ReLU}{rectified linear unit}
\newacronym{sgd}{SGD}{stochastic gradient descent}
\newacronym{mmsece}{MMSE-CE}{MMSE channel estimation}

\definecolor{sblue}{RGB}{0,0,0}
\definecolor{sred}{RGB}{200,51,130}

\newcommand{\tr}{\mathrm{tr}}

\newcommand{\matc}[1]{\mbox{\boldmath $\mathcal{#1}$}}

\newcommand{\figref}[1]{Fig. \ref{#1}}

\def\BibTeX{{\rm B\kern-.05em{\sc i\kern-.025em b}\kern-.08em
		T\kern-.1667em\lower.7ex\hbox{E}\kern-.125emX}}
\begin{document}
	
	\title{ End-to-End Learning for Uplink MU-SIMO Joint Transmitter and Non-Coherent Receiver Design in Fading Channels}

\author{Songyan Xue, Yi Ma, and Na Yi
	\thanks{Songyan Xue was with the Institute for Communication Systems (ICS), University of Surrey. He is currently with the Huawei Technologies Co., Ltd., Shanghai, China. E-mail: xuesongyan@huawei.com.}%
	\thanks{Yi Ma and Na Yi are with the Institute for Communication Systems (ICS), University of Surrey, Guildford, England, GU2 7XH. E-mail: (y.ma, n.yi)@surrey.ac.uk. Tel.: +44 1483 683609.}}%
	
\markboth{}%
{}
\maketitle

\begin{abstract}
In this paper, a novel end-to-end learning approach, namely JTRD-Net, is proposed for uplink multiuser single-input multiple-output (MU-SIMO) joint transmitter and non-coherent receiver design (JTRD) in fading channels. The basic idea lies in the use of artificial neural networks (ANNs) to replace traditional communication modules at both transmitter and receiver sides. More specifically, the transmitter side is modeled as a group of parallel linear layers, which are responsible for multiuser waveform design; and the non-coherent receiver is formed by a deep feed-forward neural network (DFNN) so as to provide multiuser detection (MUD) capabilities. The entire JTRD-Net can be trained from end to end to adapt to channel statistics through deep learning. After training, JTRD-Net can work efficiently in a non-coherent manner without requiring any levels of channel state information (CSI). In addition to the network architecture, a novel weight-initialization method, namely symmetrical-interval initialization, is proposed for JTRD-Net. It is shown that the symmetrical-interval initialization outperforms the conventional method (e.g. Xavier initialization) in terms of well-balanced convergence-rate among users. Simulation results show that the proposed JTRD-Net approach takes significant advantages in terms of reliability and {\color{sblue} scalability} over baseline schemes on both i.i.d. complex Gaussian channels and spatially-correlated channels.
\end{abstract}


\section{Introduction}
\IEEEPARstart{M}{ultiple-input multiple-output} (MIMO) technology can significantly improve the system capacity, spectral efficiency and link reliability by exploiting the spatial-domain degrees of freedom \cite{7244171}. Due to these advantages, MIMO technology has been applied in a wide range of wireless communication standards including IEEE 802.11n, \gls{lte}, and 5G \gls{nr} \cite{7084401}. To take full advantage of the spatial multiplexing-gain, most approaches require accurate channel estimation at the base station, which can be carried out by periodically transmitting pilot sequences \cite{1359139}. However, the channel estimation procedure can introduce considerable latency and training overhead. The training overhead scales linearly with the number of user terminals (UTs) \cite{1193803}. Moreover, the maximum number of UTs served in the multiuser system is limited by the number of orthogonal pilot sequences. This restriction on the availability of orthogonal resources forces the reuse of pilots for UTs in different cells. One of the main consequences of pilot-signal reuse is the pilot contamination, which has become a fundamental performance bottleneck in large-scale MIMO systems. A good overview of schemes to tackle this problem can be found in \cite{7339665} and references therein, including pre-coding, semi-blind, and blind estimation methods.
 
As an alternative, non-coherent MIMO communication systems have {\color{sblue} attracted} great attention \cite{6780655,978730,4359531,4787623}, since they require no prior knowledge of instantaneous \gls{csi} at either transmitter or receiver side. In this case, the common practice is to form the transmit signal in a way that permits accurate detection in the presence of channel uncertainty, e.g., differential encoding \cite{887864}, {\color{sblue} space-time codes} \cite{1192168,825818}. The former provides unambiguous signal reception by using modulation schemes (e.g. phase-shift-keying (PSK)) to ensure time-domain dependence of the transmitted signal. The latter is shown to be capacity-achieving in high \gls{snr} regime for block Rayleigh fading channels \cite{978730,6415385}. The idea is to carry the information in a subspace of the transmitted signal block which makes it invariant to the channel matrix multiplication, and such a subspace belongs to the Grassmannian manifold \cite{978730}. However, its computational complexity grows exponentially with the size of the decision region, which makes it unrealistic in real practice. To achieve the best performance-complexity trade-off, a group of sub-optimal detection algorithms has been proposed for Grassmannian constellations \cite{868472,4350316,8938133}. 

Owning to its powerful data learning capability, deep learning technology has achieved significant success in {\color{sblue} wide range of fields} of applications, including natural language processing, image processing, computer vision, and many others. Recently, it has been applied to wireless communication physical layer, such as signal detection \cite{8761999,9018199,8642915,9103314}, channel coding \cite{8815549,NIPS2019_8543,7926071}, and channel estimation \cite{8949757,8813060,8052521}. A relatively comprehensive survey of deep learning techniques for wireless communication systems can be found in \cite{8666641}. More importantly, deep learning enables a low-complexity end-to-end optimization of communications systems. The idea is to represent transmitter and receiver as neural networks and interpret the whole system as an autoencoder, which can be trained in a supervised manner using learning algorithms \cite{8054694}. Compared with traditional {\color{sblue} block-level system design, the end-to-end learning of a communication system is more likely and much easier to ascertain global optimality particularly in complicated communication scenarios \cite{8985539}. This is because the individual blocks therein are separately designed and optimized with different assumptions and objectives. Also, such a design highly relies on the mathematical model of wireless channel, which cannot always correctly or accurately reflect the actual propagation scenario, thereby compromising the system performance.} Besides, {\color{sblue} it has been shown that} a learned communications system can work efficiently without requiring any levels of CSI at both transmitter and receiver sides \cite{8054694,8792076,8644250}. {\color{sblue} Specifically for MIMO systems},  in \cite{2019arXiv190303711W}, the authors proposed a joint modulation and signal detection approach for {\color{sblue} single-user} MIMO system. In \cite{8437142}, deep learning-based \gls{musimo} joint transmitter and receiver design outperforms the \gls{mmse} receiver in small-size \gls{musimo} systems. In \cite{9036067}, an autoencoder-based approach outperforms the orthogonal frequency division multiplexing with index modulation (OFDM-IM) in energy-based \gls{musimo} systems.

Despite their advantages, current solutions are still challenged by the signal processing scalability with respect to the size of \gls{musimo} networks. It has been shown that most of the existing solutions can only work efficiently for special cases such as MU-SIMO with small size (e.g. $2\times 4$) and low data rate \cite{8437142,2019arXiv190303711W}. Our preliminary work in \cite{pimrc2020} has explored the potential of applying end-to-end learning in large-size \gls{musimo} systems. Moreover, current solutions only consider simple channel models (i.i.d. complex Gaussian channel). To the best of our knowledge, none of the existing works has investigated the impact of channel correlation on the end-to-end learning in \gls{musimo} systems.

Motivate by the above observations, we explore the feasibility of providing a scalable and robust solution for uplink \gls{musimo} joint transmitter and {\color{sblue} non-coherent receiver design}. Major contributions of this paper include:  
\begin{itemize}
\item The development of a novel end-to-end learning approach for uplink \gls{musimo} systems, namely JTRD-Net. In JTRD-Net, transmitters are modeled as a group of parallel linear layers, which are responsible for multiuser waveform design; and the non-coherent receiver is formed by a deep feed-forward neural network (DFNN) so as to provide multiuser detection (MUD) capabilities. The entire JTRD-Net can be trained from end to end through deep learning. After training, JTRD-Net can work efficiently in a non-coherent manner. Simulation results show that JTRD-Net outperforms baseline schemes on both i.i.d. complex Gaussian channels and spatially-correlated channels. More interestingly, it is shown that channel correlation benefits end-to-end learning in terms of reliability and training complexity. 
\item The analysis of computational complexity for the proposed JTRD-Net. It is shown that JTRD-Net has a simple network architecture with only feed-forward neural networks. Therefore, the computational complexity is mainly dominated by matrix multiplication, i.e., it bypasses matrix inversions or factorizations which are needed for most of the conventional non-coherent detection approaches. 
\item The development of a novel weight initialization method for the proposed JTRD-Net, namely symmetrical-interval weight initialization. It is shown that the symmetrical-interval initialization outperforms the conventional method (e.g., Xavier initialization) in terms of well-balanced convergence-rate among different UTs. 
\end{itemize}

The rest of this paper is organized as follows. Section II presents the system model and preliminaries. Section III presents the novel JTRD-Net approach with detailed training procedure and complexity analysis. Simulation results are presented in Section IV; and finally, Section V draws the conclusion.

\textit{Notation}: Regular letter, lower-case bold letter, and capital bold letter represent scalar, vector, and matrix, respectively. $\mathbb{R}$ represents the real field and $\mathbb{C}$ represents the complex field. $\Re(\cdot)$ and $\Im(\cdot)$ represent the real and imaginary parts of a complex number, respectively. The superscripts $(\cdot)^T$, $(\cdot)^H$, and $(\cdot)^{-1}$ represent the transpose, Hermitian, and inverse of a vector/matrix, respectively. $\left | \cdot \right |$, $\left \| \cdot \right \|$ represent the absolute value, and the $\ell_2$-norm, respectively. $\mathbb{E}\left [ \cdot \right ]$, $p(\cdot)$, $\det(\cdot)$, and $\log(\cdot)$ represent the expectation, the probability, the determinant and the logarithm function, respectively. $\tr\left\{\mathbf{A} \right\}$ is the trace of matrix $\mathbf{A}$. The operator $\mathrm{vec}\{\mathbf{A}\}$ stacks all columns of the matrix $\mathbf{A}$ on top of each other, from left to right. The notation $\left \|\mathbf{z}\right \|_{\mathbf{A}}^2$ denotes the operation of $\mathbf{z}^H\mathbf{A}\mathbf{z}$.

\section{System Model and Preliminaries}
\subsection{\gls{musimo} Uplink System Model}
Consider \gls{musimo} uplink communications, where $M$ UTs simultaneously communicate to an uplink \gls{ap} with $N$ receive antennas ($N \geq M$). The \gls{musimo} channel is assumed to be block-fading, i.e., the channel remains constant with a coherent block of length $T>1$ and changes independently between blocks, and each UT employs a single transmit-antenna to send a fixed number of data streams. Within a coherent block, each UT sends a signal vector $\mathbf{x}_m$, and the received signal at the \gls{ap} is described by the following matrix form
\begin{equation}\label{eq01}
\mathbf{Y} = \mathbf{X}\mathbf{H}+\mathbf{V}
\end{equation}
where $\mathbf{Y}\in \mathbb{C}^{T\times N}$ is the received signal block over $T$ coherent intervals, $\mathbf{X}\triangleq\left[\mathbf{x}_1,\mathbf{x}_2,\dots,\mathbf{x}_{M}\right] \in \mathbb{C}^{T\times M}$ is the transmitted signal block, $\mathbf{H} \in \mathbb{C}^{M\times N}$ is the \gls{musimo} channel matrix, $\mathbf{V}\in\mathbb{C}^{T\times N}$ is the matrix of \gls{awgn}.

We will work under the following assumptions:
\begin{enumerate}
	\item The \gls{musimo} channel matrix $\mathbf{H}$ is unknown at both transmitter and receiver, and no stochastic model is assumed for it.
	\item The transmitted signal block $\mathbf{X}$ is randomly drawn from a finite alphabet set $\matc{\mathcal{A}}=\{\mathbf{X}_1,\mathbf{X}_2,...,\mathbf{X}_{K}\}$, with equal probability. Let $\mathbf{x}_m \in \mathbb{C}^{T\times 1}$ be the $m^\mathrm{th}$ column of $\mathbf{X}$, the average transmit power of each UT is regularized by $\mathbb{E}\left [\mathbf{x}_m^H\mathbf{x}_m\right ] \leq \alpha_m P$, where $\alpha_m \geq 0$ satisfy $\sum_{m=1}^{M}\alpha_m=T$, and $P$ is the power budget.
	\item Each element in noise matrix $\mathbf{V}$ is independently drawn from $\mathcal{CN}(0,\sigma^2)$, where $\sigma^2$ is the noise variance, and $\mathbf{\Phi} = \mathbb{E}\left [\mathbf{V}\mathbf{V}^H\right ]$ denotes the noise covariance matrix.
\end{enumerate}

\subsection{Conventional Non-Coherent Detection and Problem Formulation}
Based on the system model and the above assumptions, non-coherent \gls{musimo} receiver faces a multiple hypothesis testing problem which can be solved by \cite{scharf1991statistical}
\begin{equation}\label{eq02}
\hat{\mathbf{X}}_k = \underset{\mathbf{X}_k \in \matc{\mathcal{A}}}{\argmin}  \left \|\mathbf{y}-\left(\mathbf{I}_N\otimes \mathbf{X}_k\right)\hat{\mathbf{h}}_k \right \|^2_{\mathbf{\Phi}^{-1}}
\end{equation} 
where 
\begin{equation}\label{eq03}
\hat{\mathbf{h}}_k = \left(\matc{\mathcal{X}}_k^H\matc{\mathcal{X}}_k\right)^{-1}\matc{\mathcal{X}}_k^H\mathbf{\Phi}^{-1/2}\mathbf{y}
\end{equation}
and
\begin{equation}\label{eq04}
\matc{\mathcal{X}}_k = \mathbf{\Phi}^{-1/2} \left(\mathbf{I}_N \otimes \mathbf{X}_k\right)
\end{equation}
where $\mathbf{y}=\mathrm{vec}\{\mathbf{Y}\}$, $\mathbf{I}_N$ stands for the $(N)\times(N)$ {\color{sblue} identity} matrix, {\color{sblue}and $\mathbf{A}\otimes \mathbf{B}$ for the Kronecker product between $\mathbf{A}$ and $\mathbf{B}$}. The detection can be implemented by employing $K$ parallel processors with each process corresponds to a specific codeword $\mathbf{X}_k,~_{1\leq k\leq K}$, and computes the likelihood ratio of each codeword. In order to achieve {\color{sblue} more accurate estimates} of the transmitted signal block, the codebook $\matc{\mathcal{A}}$ needs to be carefully designed. We first consider the simplest case that the codebook only have two codewords (e.g. $\mathbf{X}_i$ and $\mathbf{X}_j$, $_{i\neq j}$). Let $\mathcal{P}(\mathbf{X}_i\rightarrow \mathbf{X}_j)$ be the probability of misclassification in noiseless case, i.e., the detector mistakenly deciding $\mathbf{X}_j$ when $\mathbf{X}_i$ is transmitted, as
\begin{equation}\label{eq05}
\mathcal{P}(\mathbf{X}_i\rightarrow \mathbf{X}_j)= \mathcal{P}\left( \left \| \mathbf{d}_i \right \|^2 > \left \| \mathbf{d}_j \right \|^2\right)
\end{equation}
where
\begin{align}\label{eq06}
 \color{sblue} \mathbf{d}_{u} & \color{sblue}= \mathbf{\Phi}^{-1/2} \left(\mathbf{y}-\left(\mathbf{I}_N\otimes \mathbf{X}_{u}\right)\hat{\mathbf{h}}_{u}\right) \nonumber \\
&\color{sblue} =\mathbf{\Phi}^{-1/2}\mathbf{y} - \matc{\mathcal{X}}_{u}\hat{\mathbf{h}}_{u}, \quad \text{for} ~u = i,j
\end{align}
Since $\mathbf{X}_i$ is transmitted, then we have 
\begin{equation}\label{eq07}
\mathbf{y} = \left(\mathbf{I}_N\otimes \mathbf{X}_i\right)\mathbf{h} + \mathbf{v}
\end{equation}
where $\mathbf{h} =\mathrm{vec}\{\mathbf{H}\}$, and $\mathbf{v} =\mathrm{vec}\{\mathbf{V}\}$. Plugging \eqref{eq06} and \eqref{eq07} in \eqref{eq05} yields \cite{4359531}
\begin{equation}\label{eq08}
\mathcal{P}(\mathbf{X}_i\rightarrow \mathbf{X}_j)= \mathcal{P}\Big(\mathbf{w}^H\left(\mathbf{\Gamma}_i-\mathbf{\Gamma}_j\right)\mathbf{w} 
-2\Re\left(\mathbf{w}^H\mathbf{\Gamma}_j\matc{\mathcal{X}}_i\mathbf{h}\right)>\lambda\Big)
\end{equation}
where 
\begin{equation}\label{eq09}
\mathbf{w} = \mathbf{\Phi}^{-1/2}\mathbf{v}
\end{equation}
denotes the zero-mean white Gaussian noise, and 
\begin{equation}\label{eq10}
{\color{sblue} \mathbf{\Gamma_u} = \mathbf{I}_{NT}-\matc{\mathcal{X}}_u(\matc{\mathcal{X}}_u^H\matc{\mathcal{X}}_u)^{-1}\matc{\mathcal{X}}_u^H, \quad \text{for} ~u = i,j}
\end{equation}
is the orthogonal projector onto the orthogonal complement of the column space of $\matc{\mathcal{X}}_{i(j)}$, and
\begin{equation}\label{eq11}
\lambda = \mathbf{h}^H\matc{\mathcal{X}}_i^H\mathbf{\Gamma}_j\matc{\mathcal{X}}_i\mathbf{h}.
\end{equation}
The probability in \eqref{eq08} cannot be easily calculated. {\color{sblue} Consider the operation at high-SNR regime, and the quadratic term of $\mathbf{w}$ is negligible \cite{larsson_stoica_2003}}. Therefore, we have the following approximation
\begin{align}\label{eq12}
\mathcal{P}(\mathbf{X}_i\rightarrow \mathbf{X}_j)&\approx \mathcal{P} \Big(-2\Re\left(\mathbf{w}^H\mathbf{\Gamma}_j\matc{\mathcal{X}}_i\mathbf{h}\right)>\lambda\Big) \nonumber \\
&\approx \mathcal{Q}\Big(\frac{1}{\sqrt{2}}\sqrt{\mathbf{h}^H\mathbf{L}_{ij}\mathbf{h}}\Big)
\end{align}
where 
\begin{equation}\label{eq13}
\mathbf{L}_{ij} = \matc{\mathcal{X}}_i^H \Big(\mathbf{I}_{NT}-\matc{\mathcal{X}}_j\left(\matc{\mathcal{X}}_j^H\matc{\mathcal{X}}_j\right)^{-1}\matc{\mathcal{X}}_j^H\Big)\matc{\mathcal{X}}_i
\end{equation}
and $\mathcal{Q}(\cdot)$ is the $\mathcal{Q}$-function. \eqref{eq12} shows that the probability of misclassification depends on the channel realization $\mathbf{h}$ and the relative geometry of the codewords $\matc{\mathcal{X}}_i$ and $\matc{\mathcal{X}}_j$. Since the $\mathcal{Q}$-function is monotonically non-increasing, using the inequality \cite{4359531}
\begin{equation}\label{eq14}
\color{sblue} \mathbf{h}^H\mathbf{L}_{ij}\mathbf{h} \geq \lambda_{\mathrm{min},\mathbf{L}_{ij}}\left \| \mathbf{h} \right \|^2
\end{equation}
{\color{sblue} where $\lambda_{\mathrm{min},\mathbf{L}_{ij}}$ denotes the minimum eigenvalue of the Hermitian matrix $\mathbf{L}_{ij}$.} We can obtain the upper bound at high-SNR regime
\begin{equation}\label{eq15}
\color{sblue} \mathcal{P}(\mathbf{X}_i\rightarrow \mathbf{X}_j) \leq \mathcal{Q}\Big(\frac{1}{\sqrt{2}}\left \| \mathbf{h} \right \|\sqrt{\lambda_{\mathrm{min},\mathbf{L}_{ij}}}\Big)
\end{equation}
From \eqref{eq15} we know that the \gls{pep} is decided by the fading channel $\mathbf{h}$ as well as $\mathbf{L}_{ij}$, where the latter is directly decided by the codewords. In order to minimize the \gls{pep}, \eqref{eq15} shows that we can either {\color{sblue} maximize $\left \| \mathbf{h} \right \|$ or $\lambda_{\mathrm{min},\mathbf{L}_{ij}}$.} Since we cannot control the {\color{sblue}gain} of wireless channel $\mathbf{h}$, the only solution is to design a codebook that {\color{sblue}maximizing $\lambda_{\mathrm{min},\mathbf{L}_{ij}}$.} However, such a joint codebook optimization problem is mathematically challenging since it is a high-dimensional and non-linear problem \cite{scharf1991statistical}. Nevertheless, the optimization in \eqref{eq15} is a sub-optimal solution in nature compared with the \gls{mlsd} algorithm. Motivated by the above facts, we fundamentally rethink the joint transmit and non-coherent receiver optimization problem in \gls{musimo} systems, and consider deep learning technique as a potential solution. {\color{sblue}This is because: 1) deep learning has demonstrated remarkable performance in end-to-end design of point-to-point communications \cite{8054694}; 2) deep learning based MUD can achieve near-optimal performance in various scenarios \cite{8761999,9018199,8642915,9103314}; 3) most of the deep learning algorithms have parallel computing architecture in nature, which means the computational complexity can be well handled by employing the high-performance computing technologies, e.g., Graphical processing unit (GPU) and field programmable gate array (FPGA).} 
 
\section{End-to-End Learning for \gls{musimo} System}
In this section, we first introduce the proposed end-to-end learning approach for uplink \gls{musimo} joint transmitter and non-coherent receiver design, namely JTRD-Net. Afterward, the training procedure and computational complexity analysis are provided.
\subsection{JTRD-Net Architecture}
\begin{figure*}[!t]
	\centering
	\includegraphics[width=6.5in]{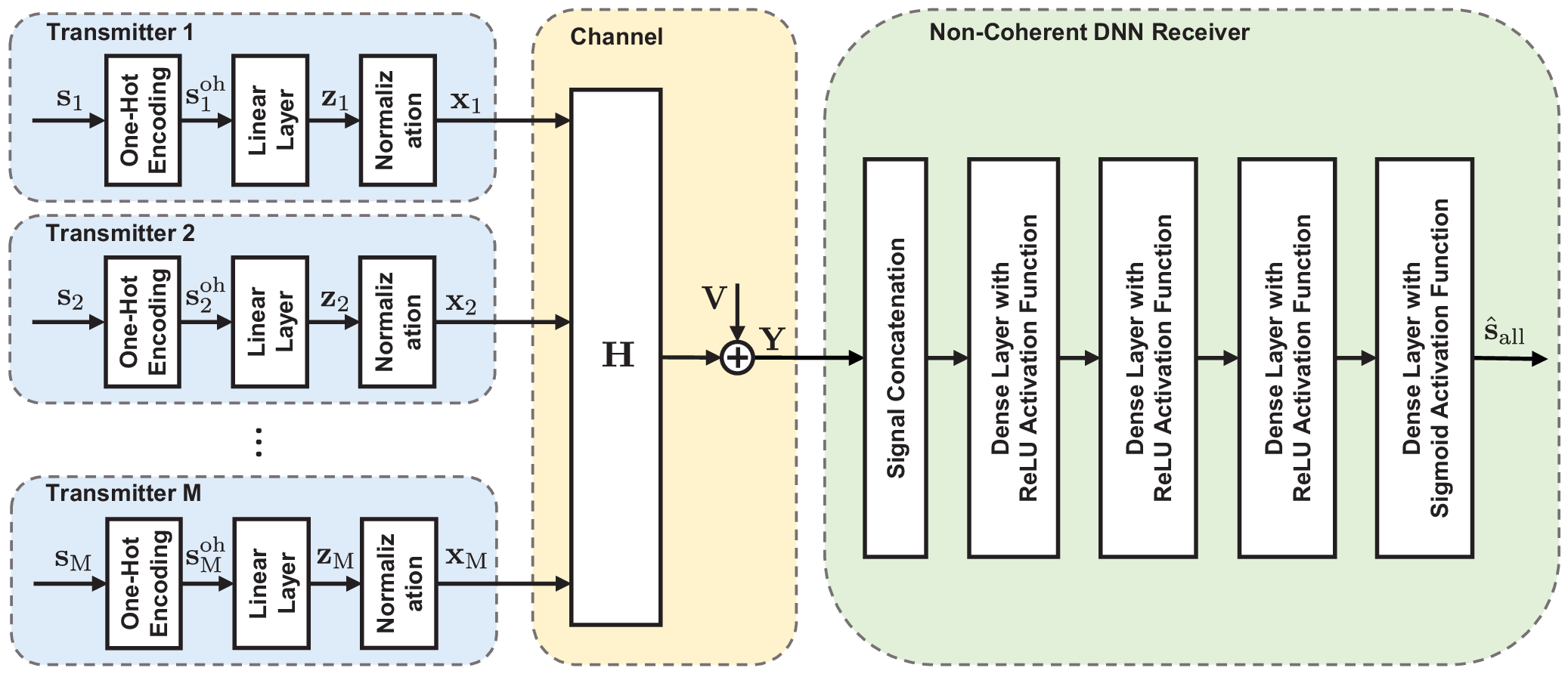}
	\caption{Block diagram of the proposed JTRD-Net approach.}
	\label{figure1}
\end{figure*}
\figref{figure1} illustrates the block diagram of the proposed JTRD-Net approach for uplink \gls{musimo} joint transmitter and non-coherent receiver design. In JTRD-Net, both multiuser transmitters and non-coherent receiver are designed by using neural networks. Specifically, the transmitter is modeled as a group of parallel linear layers with each followed by a normalization function. The input to the linear layer is a one-hot vector $\mathbf{s}_m^{\mathrm{oh}}\in \mathbb{R}^{(L)\times(1)},~_{1\leq m\leq M}$, which is obtained by one-hot encoding \cite{RODRIGUEZ201821} the binary information-bearing vector $\mathbf{s}_m\in \mathbb{R}^{(J)\times(1)},~_{1\leq m\leq M}$, where $J=\log_2L$ is the number of information bits per codeword. For instance, two information-bits can be encoded as
\begin{equation}
 00 \rightarrow 0001, 01 \rightarrow 0010, 10 \rightarrow 0100, 11 \rightarrow 1000 \nonumber
\end{equation}
The output of the linear layer is $\mathbf{z}_m=\mathbf{W}_m\mathbf{s}^{\mathrm{oh}}_m,~_{1\leq m\leq M}$, where $\mathbf{W}_m$ stands for the weighting matrix. {\color{sblue} Here the linear layer does not need a bias node, because the weighting matrix can be regarded as codebook independently. Adding a bias node is equivalent to add the same vector to each column of the codebook, which does not make any sense in joint codebook design.} Besides, the one-hot vector can be viewed as a codeword selector, which picks one column from $\mathbf{W}_m$ according to the information-bearing vector $\mathbf{s}_m$; and $\mathbf{W}_m$ is the user-specific codebook. It is worth noting that most of the existing deep learning algorithms are based on real-valued operations, but the wireless communication systems are normally modeled as complex-valued symbols \cite{8437142} (e.g. constellations and channel coefficients). To facilitate the learning and communication procedure, it is common practice (see \cite{8761999,9018199,8642915,9103314,8054694}) to convert complex signals to their real signal equivalent version by \footnote{In the rest of this paper, we do not use the doubled size for the sake of mathematical notation simplicity.}
\begin{equation}\label{eq16}
\mathbf{a}_{\mathrm{real}} = \begin{bmatrix}
\Re(\mathbf{a}) \\ 
\Im(\mathbf{a})
\end{bmatrix}
\end{equation}
Therefore, we have $\mathbf{W}_m \in \mathbb{R}^{(2T)\times(L)}$ and $\mathbf{z}_m \in \mathbb{R}^{(2T)\times(1)}$, which are equivalent to their complex-valued form $\mathbf{W}_m \in \mathbb{C}^{(T)\times(L)}$ and $\mathbf{z}_m \in \mathbb{C}^{(T)\times(1)}$, where $T$ is the coherent-block length as we introduced in \eqref{eq01}. Before transmitting through the channel, $\mathbf{z}_m$ needs to be normalized in order to meet the power constraint, which is given by
\begin{align}\label{eq17}
\mathbf{x}_m &= f_\mathrm{norm}(\mathbf{z}_m) \nonumber \\
&=\sqrt{\alpha_m P} \cdot  \frac{\mathbf{z}_m}{\sqrt{\sum \left \| \mathbf{z}_m \right \|^2}} 
\end{align}
where $\alpha_m$ and $P$ are the power-constraint parameters. The transmit signal is expressible as 
\begin{equation}\label{eq18}
\mathbf{X} = [\mathbf{x}_1, \mathbf{x}_2, \dots, \mathbf{x}_M]^T
\end{equation}
According to the system model in \eqref{eq01}, the received signal is given by
\begin{equation}\label{eq19}
\mathbf{y} = \mathbf{x}^T \cdot (\mathbf{I}_T \otimes \mathbf{H}_\mathrm{real} ) + \mathbf{v}_\mathrm{real} 
\end{equation}
where $\mathbf{x}=\mathrm{vec}\{\mathbf{X}\} \in \mathbb{R}^{(2MT)\times(1)}$, $\mathbf{v}_\mathrm{real}  \in \mathbb{R}^{(2NT)\times(1)}$ stands for the \gls{awgn} vector, and  
\begin{equation}\label{eq20}
\mathbf{H}_\mathrm{real} = \begin{bmatrix}
\Re(\mathbf{H})&-\Im(\mathbf{H})\\ 
\Im(\mathbf{H})&~~\Re(\mathbf{H})
\end{bmatrix}
\end{equation}
for the channel matrix in its real-valued equivalent form. At the receiver side, $\mathbf{y}$ is utilized as the input to the non-coherent \gls{dnn} receiver. The non-coherent \gls{dnn} receiver is modeled as a FDNN with three hidden layers. The activation function for the hidden layers is the \gls{relu}, and for the output layer is standard logistic function (Sigmoid). The final output of the JTRD-Net $\hat{\mathbf{s}}_{\mathrm{all}} \in \mathbb{R}^{(JM)\times (1)}$ is the estimate of the original information-bearing vectors.

\subsection{Training Procedure of JTRD-Net}
The entire JTRD-Net can be trained from end-to-end with the aim of minimizing the following binary cross-entropy cost function
\begin{equation}\label{eq21}
J(\matc{\varphi})= -\frac{1}{\left|\mathcal{B}\right|}\sum_{i=1}^{\left|\mathcal{B}\right|} \Big(\mathbf{s}_{\mathrm{all}}^{(i)}\log \hat{\mathbf{s}}_{\mathrm{all}}^{(i)} + (1-\mathbf{s}_{\mathrm{all}}^{(i)})\log (1-\hat{\mathbf{s}}_{\mathrm{all}}^{(i)}) \Big)
\end{equation}
where $\mathcal{B}$ stands for the training mini-batch with size of $\mathcal{B}$, the superscript for the index, and $\matc{\varphi}=\{\mathbf{W},\mathbf{b}\}$ for the trainable parameters in the JTRD-Net. Moreover, $\mathbf{s}_{\mathrm{all}} \in \mathbb{R}^{(JM)\times (1)}$ is the reference training target, which is obatined by concatenating all information-bearing vectors as
\begin{equation}\label{eq22}
\mathbf{s}_{\mathrm{all}} = [\mathbf{s}_1^T,\mathbf{s}_2^T,\dots,\mathbf{s}_M^T]^T
\end{equation}

The difference between the training of JTRD-Net and other deep learning applications is that the channel matrix $\mathbf{H}$ needs to be considered in the \gls{bp} procedure. Otherwise, the linear layers at the transmitter side can not obtain correct gradients for parameter updating. To elaborate a little further, we assume the gradient of the received signal $\mathbf{y}$ is $\nabla_\mathbf{y}J(\matc{\varphi})\in \mathbb{R}^{(2NT)\times (1)}$. \footnote{The training {\color{sblue} batch} is not considered to simplify the mathematical expressions.} Here we skip the introduction of calculating the gradient for the receiver-side hidden layers, since it can be easily obtained by utilizing standard \gls{bp} algorithm. In order to update the transmitter-side neural networks, we need to calculate the gradient for the transmitted signal block. {\color{sblue} To this end, channel matrix is assumed to be known at the training stage,} then the gradient of the transmitted-signal block is expressible as
\begin{equation}\label{eq23}
\nabla_\mathbf{x}J(\matc{\varphi}) = (\mathbf{I}_T \otimes \mathbf{H}_\mathrm{real} ) \cdot \nabla_\mathbf{y}J(\matc{\varphi})
\end{equation}
{\color{sblue} Note that the above assumption does not affect the non-coherent detection, since the channel knowledge is only utilized for backpropagation at training stage. After training, the entire JTRD-Net can still work efficiently in a non-coherent manner without requiring any levels of \gls{csi}.} The gradient of each transmitted signal-block can be obtained by reshaping $\nabla_\mathbf{x}J(\matc{\varphi}) \in \mathbb{R}^{(2MT)\times (1)}$ to a $(2T)\times(M)$ matrix in a row-major order, and the $m^\mathrm{th}$ column of the matrix is the gradient of $\mathbf{x}_m$, i.e., $\nabla_{\mathbf{x}_m}J(\matc{\varphi})\in \mathbb{R}^{(2T)\times (1)},~_{1\leq m \leq M}$. Afterwards, the gradient of the weighting matrix on each linear layer can be computed by
\begin{equation}\label{eq24}
\nabla_{\mathbf{W}_m}J(\matc{\varphi}) = \Big(\nabla_{\mathbf{x}_m}J(\matc{\varphi}) \odot {f}'_\mathrm{norm}(\mathbf{z}_m) \Big) \cdot {\mathbf{s}_m^\mathrm{oh}}^T
\end{equation}
where $\odot$ stands for point-wise multiplication, and ${f}'_\mathrm{norm}(\mathbf{z}_m)$ for the derivative of the normalization function which is given by
\begin{align}\label{eq25}
{f}'_\mathrm{norm}(\mathbf{z}_m) &= \frac{\partial}{\partial\mathbf{z}_m}\bigg(\sqrt{\alpha_m P} \cdot \frac{\mathbf{z}_m}{\sqrt{\sum \left \| \mathbf{z}_m \right \|^2}} \bigg)  \nonumber \\
&= \sqrt{\alpha_m P} \cdot \frac{\left(\mathbf{1}_{(2T,1)} \otimes \sum\left \| \mathbf{z}_m \right \|^2\right) - \mathbf{z}_m}{\left(\sum\left \| \mathbf{z}_m \right \|^2\right)^{3/2}}
\end{align}
where $\mathbf{1}_{(2T,1)}$ stands for an all-one vector with size of $(2T)\times (1) $. Moreover, the training procedure does not need to be implemented to the one-hot encoding function, because it consists of no trainable parameters and can be viewed as a bijective mapping between $\mathbf{s}_m$ and $\mathbf{s}_m^\mathrm{oh}$.

\vspace{-1em}
\subsection{Complexity Analysis}
{\color{sblue} Define $b$ as the size of the mini-batch, the computational complexity for the JTRD-Net is approximately $\mathcal{O}(b(4^{J} MT+L_{h}N^2T^2+L_{h}M))$ in the training procedure ($T\geq M$), and $\mathcal{O}(4^{J} MT+L_{h}N^2T^2+L_{h}M)$ in the communication procedure, where $J$ stands for the number of information bits per codeword per user and $L_{h}$ for the number of neurons on the receive-DNN hidden layer which might vary with the size of the MU-SIMO network.} The complexity is mainly dominated by matrix multiplications. To put this in perspective, the expectation propagation (EP) based non-coherent MU-SIMO detection normally has a computational complexity around $\mathcal{O}(M^72^Kn_\mathrm{iteration})$ \cite{8645403}, where $n_\mathrm{iteration}$ is the number of detection iterations. The maximum-likelihood detection for Grassmannian modulation has a complexity of $\mathcal{O}(2^{MK})$ dominated by an exhaustive search. {\color{sblue} It has been shown, in the literature, that most of the conventional approaches can only work efficiently when spatial-domain user load is relatively low (e.g. $M=4$). Also, the coherent block length $T$ has to be much larger than $M$ (e.g. $T > 2M$). In the simulation, we will show that the proposed JTRD-Net is scalable in term of the spatial-domain user load, and the coherent block length does not need to be that long.} Moreover, recall that the proposed JTRD-Net approach is mainly formed by neural networks, it fits into the trend of high-performance computing technologies that highly rely on parallel processing to improve the computing speed, the capacity of multi-task execution as well as the computing energy-efficiency. {\color{sblue} This is an important feature as it equips the receiver with a great potential of providing ultra-low latency and energy-efficient signal processing that is one of the key requirements for future wireless networks \cite{8636206}.}

\section{Simulation Results}
This section presents the simulation result and performance analysis. The data sets and implementation details are introduced at the beginning, followed by the introduction of the proposed weight initialization method. Afterwards, a comprehensive performance evaluation is provided, which demonstrates the performance of the proposed JTRD-Net approach.

\subsection{Data set and Implementation Details}
In traditional \gls{ai} applications including image classification and nature language processing, learning algorithms and models depend heavily on training data-set (e.g. MNIST \cite{lecun-mnisthandwrittendigit-2010}, MS-COCO \cite{lin2014microsoft}, and CIFAR-10 \cite{Krizhevsky09learningmultiple}). In many cases, it is very difficult to build training data-sets that are large enough to meet the training requirements. However, this problem can be easily solved in the wireless communication domain, since we are dealing with artificially manufactured data (e.g. modulation and coding) which can be accurately generated. Therefore, we would like to define the data generation routines instead of giving a specific training dataset in this work.

As far as supervised learning is concerned, the training data-set consists of a number of randomly generated pairs. The training input is a group of one-hot vectors $\mathbf{s}_m,~_{1\leq m\leq M}$, and the referenced training target is a binary vector $\mathbf{s}_\mathrm{all}$ as we introduced in \eqref{eq19}. For each training iteration, channel matrix is randomly generated subject to specific channel models. In this work, we considered three different channel models, including i.i.d. complex Gaussian MIMO channel, Kronecker MIMO channel, and 3GPP 3D MIMO channel. Specifically, each element of the i.i.d. complex Gaussian MIMO channel $\mathbf{H}_\mathrm{R}$ is $h_{ij}\sim \mathcal{N}\mathcal{C}(0,1/M)$, while the Kronecker MIMO channel is described by 
\begin{equation}\label{eq26}
\mathbf{H}_\mathrm{K} = \mathbf{R}_\mathrm{R}^{1/2}\mathbf{H}_\mathrm{W}^{~}\mathbf{R}_\mathrm{T}^{1/2}
\end{equation}
where $\mathbf{R}_\mathrm{T}$ and $\mathbf{R}_\mathrm{R}$ denote the transmitter and receiver side spatial-correlation matrices, respectively, and $\mathbf{H}_\mathrm{W}$ is independently and identically distributed as circular-symmetric complex Gaussian with zero-mean and unit variance, which is generated according to the exponential correlation model in \cite{951380} with the same correlation coefficient $\rho$. For the case of realistic channels, we consider 3GPP 3D MIMO channel \cite{DBLP:journals/corr/MondalTVVGNLZZL15}, and implemented in QuaDRiGa channel simulator \cite{6758357}. The \gls{ap} is assumed to be equipped with 4 dual-polarized antennas with a height of 25m. The sector coverage is $120^{\circ}$, and 4 single-antenna UTs are randomly dropped with a radius range of 500m. Users are modeled to move along a linear trajectory with a speed of 1 m/s. Besides, {\color{sblue}perfect power control is assumed, which normalizes the average received power across antennas to one.}

{\color{sblue}At each training iteration, we randomly generate a group of user-specific information-bearing bits as well as a MU-SIMO channel matrix.} In the evaluation stage, network is tested until the number of error bits reaches a certain threshold, e.g., 1,000. The network is trained by using standard \gls{bp} \cite{10.5555/201784.201785} with the mini-batch gradient descent algorithm, and the size of the mini-batch is set to be 100. To train the network well, Adam optimizer \cite{2014arXiv1412.6980K} is also utilized with an initial learning rate of 0.001. With the number of training epoch increases, we dynamically decrease the learning rate until reaching a given threshold, which can be expressed as
\begin{equation}\label{eq27}
\eta_{n_\mathrm{epoch}} = \max (\eta_i/\sqrt[4]{n_\mathrm{iteration}},\eta_\mathrm{low})
\end{equation}
where $\eta_i$ stands for the initial learning rate, $n_\mathrm{iteration}$ for the number of training epoch, and $\eta_\mathrm{low}$ for the lower bound of the learning rate, which is set to be $10^{-5}$ in this work; {\color{sblue} as we found that a smaller learning rate can achieve better convergence performance particularly at the later training stage.}

All simulations are run on a Dell PowerEdge R730 2x8-Core E5-2667v4 server, and implemented in Matlab. {\color{sblue} We release our source code at github.com/jtrdnet/source$\_$code.}

{\color{sblue}
\begin{remark}
	One of the reviewers pointed out that the simulated system considers the use of a practical but relatively small number of antennas compared to massive-MIMO systems. The training complexity and memory requirement might grow significantly with the increase of antenna size. This is indeed a very important issue for the feedforward network. A potential approach is to employ the bi-directional long short-term memory (bi-LSTM) structure to improve the training scalability; and this could be a very good research direction for future work.
\end{remark}}

\begin{figure}[!t]
	\centering
	\includegraphics[width=3.3in]{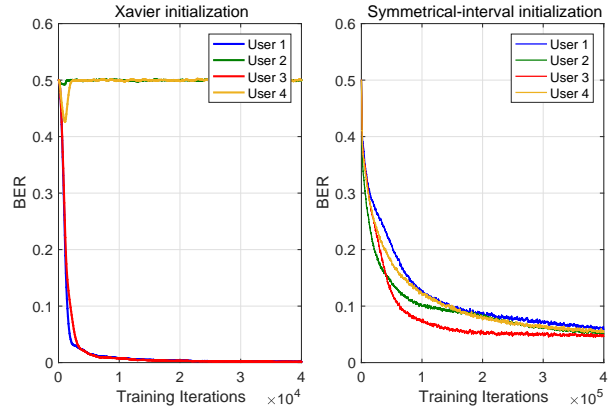}
	\caption{The convergence performance of the Xavier initialization (left) and the proposed symmetrical-interval initialization method (right).}
	\label{figure2}
\end{figure}

\subsection{Weight Initialization}
In addition to the above settings, we proposed a novel weight initialization method for the transmitter-side linear layers in the JTRD-Net. It is well known that weight initialization plays an important role in neural network training, since it directly affects the convergence performance. Traditional initialization method such as Xavier initialization \cite{GlorotAISTATS2010}, randomly generates the coefficients of weighting matrices by using the following heuristic
\begin{equation}\label{eq28}
\mathbf{W}_{ij} \sim U\Big(0,\frac{1}{\sqrt{n}}\Big)
\end{equation}
where $U(a,b)$ denotes the uniform distribution in the interval of $(a-b,a+b)$, and $n$ is the size of the input to the current layer. Although Xavier initialization has been demonstrated to achieve fast convergence performance in many deep learning applications, it is not a good solution for the proposed JTRD-Net approach. The reason is that the initialized coefficients might differ by orders of magnitude, which will result in unbalanced training among different users; as shown in \figref{figure2}-(left). To tackle this issue, we proposed a modified weight-initialization method for JTRD-Net, namely symmetrical-interval initialization, which generates the coefficients by using the following heuristic
\begin{equation}\label{eq29}
\mathbf{W}_{ij} \sim \bigg(U\Big(-\frac{1}{\sqrt{n} },\zeta\Big)\cup U\Big(\frac{1}{\sqrt{n}},\zeta \Big)\bigg)
\end{equation}
where $\zeta$ is an arbitrary number with at least one order of magnitude smaller than $1/\sqrt{n}$. By such means, the convergence among different users can be largely improved; as shown in \figref{figure2}-(right) that all users are able to achieve a balanced convergence performance. {\color{sblue} It is perhaps worth noting that the proposed method seems require much more training iterations than the conventional method. This is due to the fact that the conventional method quickly drops to a local minima, as we can see two users occupy nearly all the resources and other users fail to transmit their information.}

\subsection{Simulations and Performance Evaluation}
In this section, we study the performance of the proposed JTRD-Net approach under various channel models. The performance is evaluated using the \gls{ber} averaging over sufficient Monte-Carlo trials of block fading channels, and compared to wide range of baselines under different scenarios. The \gls{snr} of the system, defined as
\begin{equation}\label{eq30}
\mathrm{SNR} = \frac{\mathbb{E} \left \|\mathbf{H}\mathbf{x}_m  \right \|^2 }{\mathbb{E}\left \|\mathbf{v}_m  \right \|^2}
\end{equation}
is to measure the noise level.

\begin{figure}[!t]
	\centering
	\includegraphics[width=3.3in]{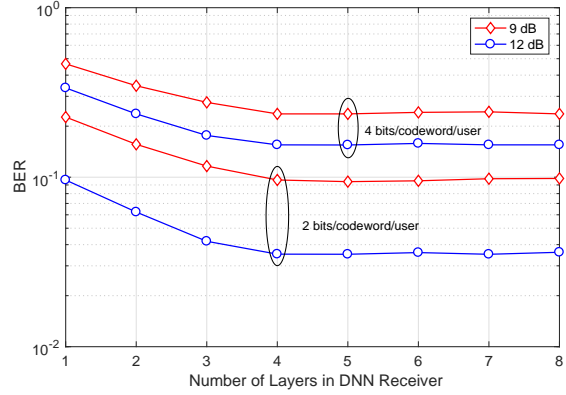}
	\caption{BER performance of the proposed JTRD-Net approach versus the number of layers in non-coherent \gls{dnn} receiver with different transmission-rates under 4-by-8 i.i.d. complex Gaussian MIMO channels.}
	\label{figure3}
\end{figure}

\begin{figure}[!t]
	\centering
	\includegraphics[width=3.3in]{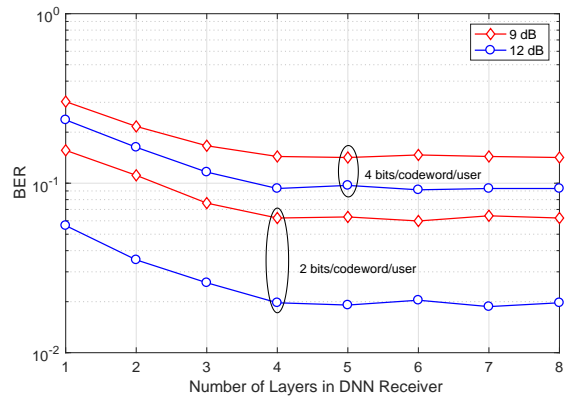}
	\caption{BER performance of the proposed JTRD-Net approach versus the number of layers in non-coherent \gls{dnn} receiver with different transmission-rates under 4-by-8 Kronecker MIMO channels.}
	\label{figure4}
\end{figure}

\subsubsection{Analysis of Network Size}
\begin{figure}[!t]
	\centering
	\includegraphics[width=3.3in]{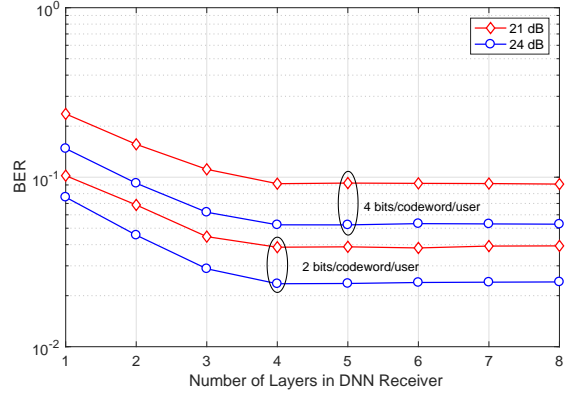}
	\caption{BER performance of the proposed JTRD-Net approach versus the number of layers in non-coherent \gls{dnn} receiver with different transmission-rates under 4-by-8 3GPP MIMO channels.}
	\label{figure4-1}
\end{figure}

 \figref{figure3}, \figref{figure4} and {\color{sblue} \figref{figure4-1} illustrate the \gls{ber} performance of the proposed JTRD-Net approach versus the number of layers in non-coherent \gls{dnn} receiver with different transmission rates (e.g. 2 bits/codeword/user and 4 bits/codeword/user) under different channel models. Besides, two training SNRs are considered (e.g. $\mathrm{SNR} =$ 9 dB and 12 dB for i.i.d complex Gaussian channels and Kronecker channels, and $\mathrm{SNR} =$ 21 dB and 24 dB for 3GPP MIMO channel); the system environment is 4-by-8  \gls{musimo}}. It is shown that JTRD-Net converges within four layers under both channel models. Further increases layers can not improve the detection performance as the \gls{ber} curve converges to a certain value. Besides, it is shown that lower transmission-rate is more sensitive to the change of \gls{snr}s, because the gap between the bottom two curves is much larger than the other one. Based on these results, the number of layers in non-coherent \gls{dnn} receiver is set to be 4 in the following simulations.

\begin{table}[!t]
	\footnotesize
	\centering
	\caption{Layout of the JTRD-Net}\label{table1}
	\renewcommand{\arraystretch}{1}
	\begin{tabular}{p{1.5cm}|p{2cm}|p{2.2cm}}
		\hline
		\textbf{JTRD-Net} &\textbf{Layer} &\textbf{Output dimension} \\ \hline
		&Input  & $L$ \\ 
		\multicolumn{1}{c|}{Transmitter}	&Dense + Linear  & $2T$ \\ 
		&Normalization  & $2T$ \\ \hline
		&Concatenation  & $2NT$ \\ 
		&Dense + ReLU & $1024$   \\ 
		\multicolumn{1}{c|}{Receiver}&Dense + ReLU & $512$   \\
		&Dense + ReLU & $256$   \\
		&Dense + Sigmoid & $JM$   \\ \hline
	\end{tabular}
\end{table} 
\subsubsection{Analysis of Coherent Block Length}
\begin{figure}[!t]
 	\centering
 	\includegraphics[width=3.3in]{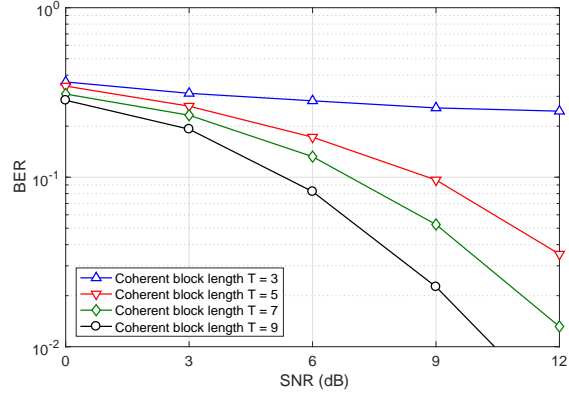}
 	\caption{Average BER performance of the proposed JTRD-Net approach versus the length of coherent block under 4-by-4 i.i.d. complex Gaussian MIMO channels.}
 	\label{figure5}
 \end{figure}

\figref{figure5} illustrate the average \gls{ber} performance of the proposed JTRD-Net approach versus the length of coherent block under 4-by-4 i.i.d. complex Gaussian MIMO channels. The aim is to investigate the effects of coherent-block length on the detection performance. It is shown that the detection performance increases with the length of the coherent block. This phenomenon is easy to understand since the time-domain degree of freedom introduces power gain to the signal detection. It is worth noting that the pilot-based channel estimation approach requires at least $T=M=4$ time slots to estimate channel. Therefore, the minimum coherent-block length for pilot-based solutions is $T=M+1$. To facilitate the performance comparison in the following simulations, we set the length of coherent block to be $M+1$.

\subsubsection{i.i.d. complex Gaussian Channels}
This section aims to investigate the \gls{ber} performance of the proposed JTRD-Net approach with different system configurations under i.i.d. complex Gaussian MIMO channels. {\color{sblue} Due to the lack of comparable algorithms, the performance of the proposed JTRD-Net approach is compared with pilot-based channel estimation approaches, e.g., \gls{mmsece}  with MMSE equalization or \gls{mlsd} algorithm, and a conventional hand-engineered non-coherent detection approach, e.g., POCIS \cite{8645403}.} {\color{sblue} The layout of the JTRD-Net is listed in Table \ref{table1}, which is obtained by trying different combinations; and the above setup is believed to offer the best training performance.}
 
 \begin{figure}[!t]
 	\centering
 	\includegraphics[width=3.3in]{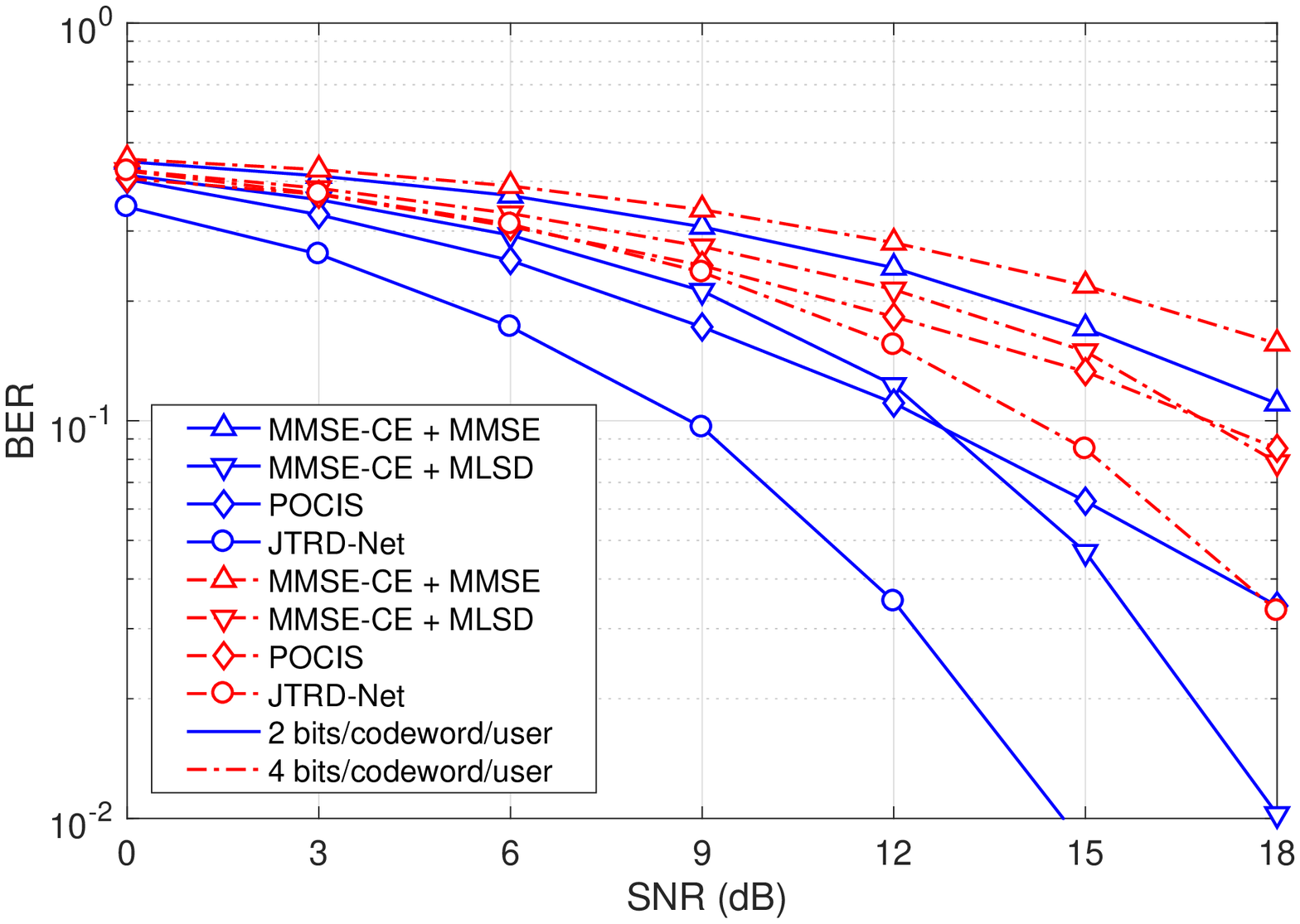}
 	\caption{Average BER performance comparison of the proposed JTRD-Net approach versus other MIMO detection algorithms with different transmission-rates under 4-by-4 i.i.d. complex Gaussian MIMO channels.}
 	\label{figure6}
 \end{figure}

 \begin{figure}[!t]
 	\centering
 	\includegraphics[width=3.3in]{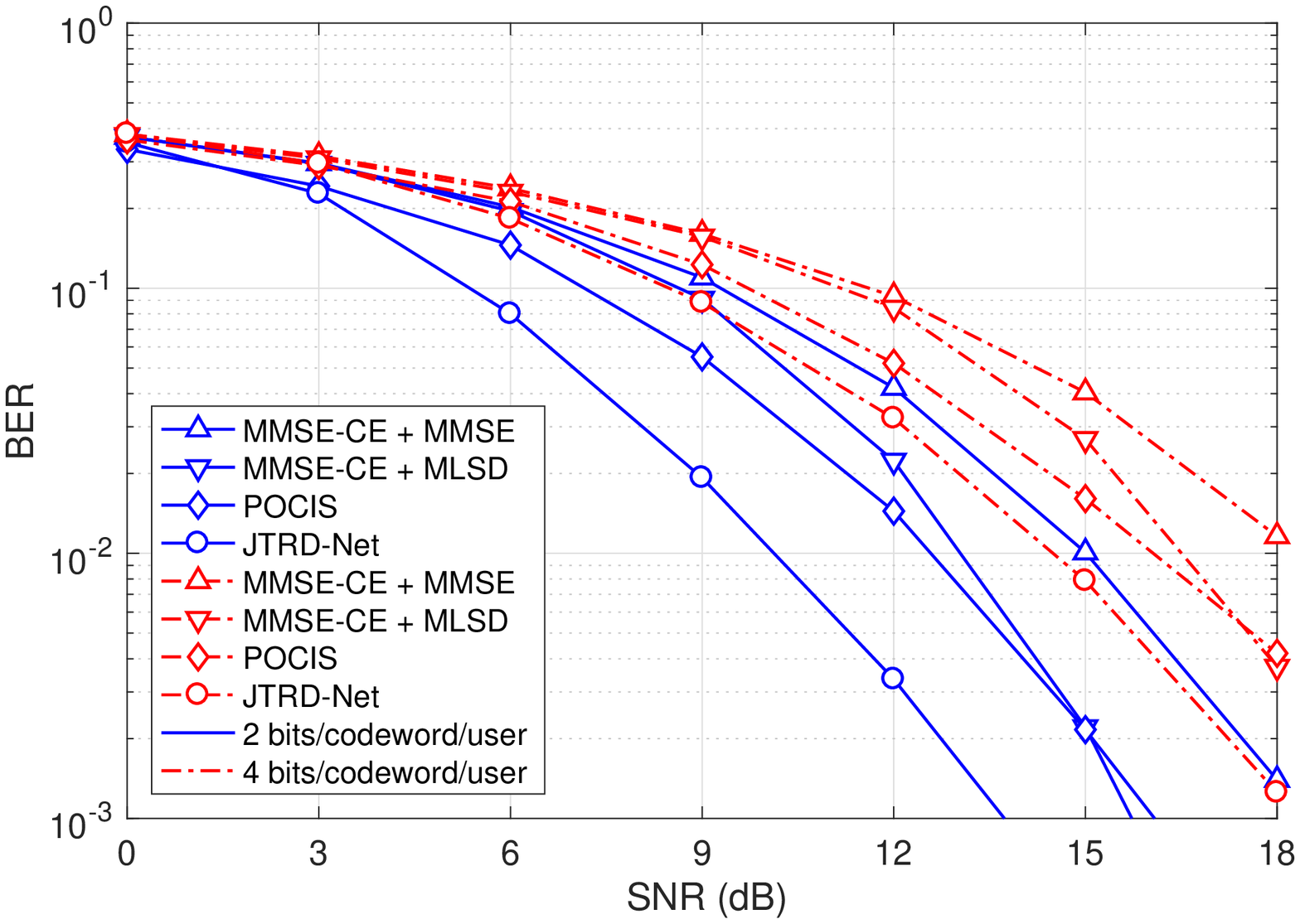}
 	\caption{Average BER performance comparison of the proposed JTRD-Net approach versus other MIMO detection algorithms with different transmission-rates under 4-by-8 i.i.d. complex Gaussian MIMO channels.}
 	\label{figure7}
 \end{figure}
 
 \begin{figure}[!t]
 	\centering
 	\includegraphics[width=3.3in]{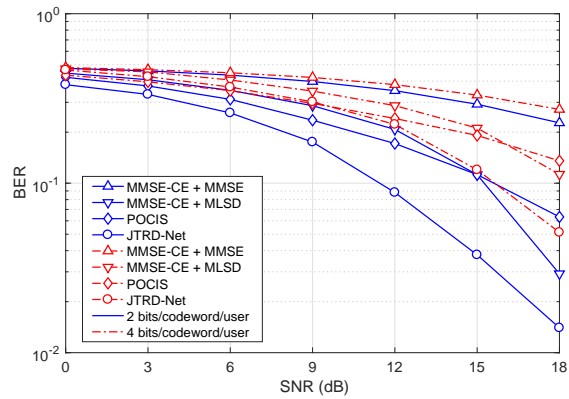}
 	\caption{Average BER performance comparison of the proposed JTRD-Net approach versus other MIMO detection algorithms with different transmission-rates under 8-by-8 i.i.d. complex Gaussian MIMO channels.}
 	\label{figure8}
 \end{figure}

\figref{figure6} illustrates the \gls{ber} performance of the JTRD-Net approach with different data rates under 4-by-4 i.i.d. complex Gaussian MIMO channels. The network is trained at \gls{snr} of 12 dB in 2 bits/codeword/user case and 15 dB in 4 bits/codeword/user case. For 2 bits/codeword/user, it is shown that the POCIS detector slightly outperforms the \gls{mmsece} with \gls{mlsd} at low \gls{snr} regime. The proposed JTRD-Net approach outperforms the baseline schemes for at least 3.2 dB at high \gls{snr} regime. The gain mainly comes from the joint transmitter and receiver optimization process. For 4 bits/codeword/user, similar phenomenons have been observed. The performance gap between the POCIS detector and the \gls{mmsece} with \gls{mlsd} increases to approximately 4 dB at high SNR. Meanwhile, the proposed JTRD-Net approach still largely outperforms conventional baselines. but the performance gap between JTRD-Net and POCIS is reduced to approximately 3 dB at high \gls{snr}. This is potentially due to the increasing data rate introduces additional training complexity for joint waveform design, i.e., the decision region has been largely increased.

\figref{figure7} illustrates the \gls{ber} performance of the JTRD-Net approach with different data rates under 4-by-8 i.i.d. complex Gaussian MIMO channels. The network is trained at \gls{snr} of 12 dB in 2 bits/codeword/user case and 15 dB in 4 bits/codeword/user case. Benefited by the spatial-domain diversity gain, all detection algorithms have their performance improved compared with the 4-by-4 \gls{musimo} system. For 2 bits/codeword/user, the gap between JTRD-Net and the POCIS detector is approximately 2.2 dB at \gls{ber} of $10^{-3}$, and JTRD-Net largely outperforms pilot-based approaches throughout the whole \gls{snr} range. For 4 bits/codeword/user, the performance gap between JTRD-Net and POCIS is slightly reduced with approximately 2 dB at \gls{ber} of $10^{-3}$. The performance degradation is caused by the expansion of the decision region. Moreover, JTRD-Net fails to achieve the full diversity order at high \gls{snr} due to the {\color{sblue} channel learning imperfection, i.e., channel randomness.}

\figref{figure8} illustrates the \gls{ber} performance of the JTRD-Net approach with different data rates under 8-by-8 i.i.d. complex Gaussian MIMO channels. {\color{sblue} This experiment is designed to investigate the scalability of the JTRD-Net, since the learning difficulty mainly lies in the user codebook design. Thus, 8 user case is expected to be much more challenging compared with 4 user case.} The network is trained at \gls{snr} of 15 dB in 2 bits/codeword/user case and 18 dB in 4 bits/codeword/user case. With the increasing spatial-domain user load, all of the detection algorithms have their performance degraded compared with the previous case. For 2 bits/codeword/user, the performance improvement from pilot-based scheme to the JTRD-Net is approximately 2 dB at high \gls{snr}, and JTRD-Net largely outperforms the POCIS detector throughout the whole SNR range. For 4 bits/codeword/user, similar phenomenons have been observed. The performance gap between JTRD-Net and POCIS is approximately 4 dB at high \gls{snr}, and JTRD-Net outperforms pilot-based schemes for at least 3 dB at high \gls{snr}.

\begin{figure}[!t]
	\centering
	\includegraphics[width=3.3in]{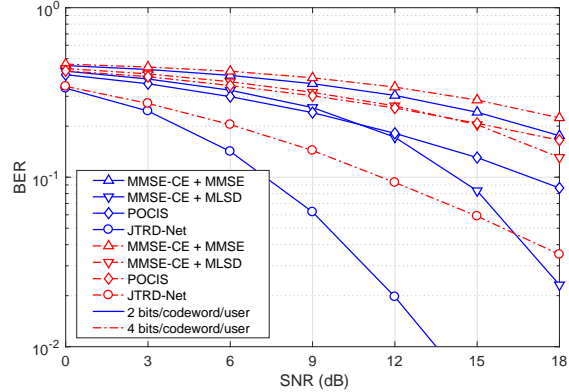}
	\caption{Average BER performance comparison of the proposed JTRD-Net approach versus other MIMO detection algorithms with different transmission rates under 4-by-4 Kronecker MIMO channels.}
	\label{figure9}
\end{figure}

\begin{figure}[!t]
	\centering
	\includegraphics[width=3.3in]{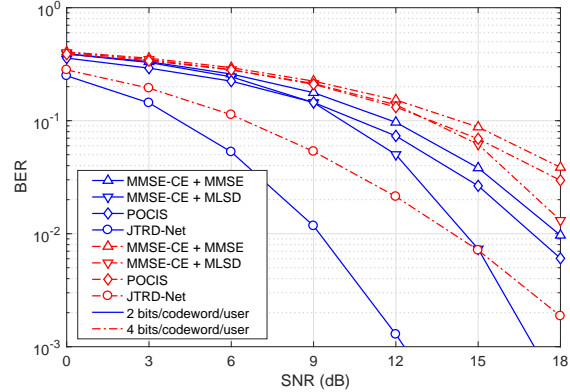}
	\caption{Average BER performance comparison of the proposed JTRD-Net approach versus other MIMO detection algorithms with different transmission rates under 4-by-8 Kronecker MIMO channels.}
	\label{figure10}
\end{figure}

\subsubsection{Correlated MIMO Channel}
This section aims to investigate the \gls{ber} performance of the proposed JTRD-Net approach with different system configurations under correlated MIMO channels (e.g. Kronecker MIMO channels with $\rho = 0.5$). All of the network configurations and baselines remain unchanged as we utilized under i.i.d. complex Gaussian MIMO channels.

\figref{figure9} illustrates the \gls{ber} performance of the JTRD-Net approach with different data rates under 4-by-4 Kronecker MIMO channels. The network is trained at \gls{snr} of 12 dB in 2 bits/codeword/user case and 15 dB in 4 bits/codeword/user case. In this figure, it is shown that the conventional detection algorithms have their performance degraded for approximately 5 dB compared with the \gls{ber} under i.i.d. complex Gaussian MIMO channels. Meanwhile, it is also shown that the detection accuracy of the proposed JTRD-Net approach is increased for around 1.3 dB at high \gls{snr}. Through our analysis of the neural-network designed codebooks, we found that the transmission power is jointly optimized among different users' codebooks, which can largely mitigate the inter-user interference (IUI). {\color{sblue} This means that the transmit power is not evenly distributed over $T$ coherent block length, and is indeed jointly optimized by backpropagation algorithm throughout the training procedure.} For 2 bits/codeword/user, the POCIS detector achieves nearly the same performance as the \gls{mmsece} with \gls{mlsd} algorithm at low \gls{snr}. The performance gap between the JTRD-Net approach and the pilot-based schemes is approximately 6.5 dB at \gls{ber} of $10^{-2}$. For 4 bits/codeword/user, this gap is further increased to approximately 9 dB at high \gls{snr}. Again, the POCIS detector achieves nearly the same performance as the pilot-based solutions. 

\begin{figure}[!t]
	\centering
	\includegraphics[width=3.3in]{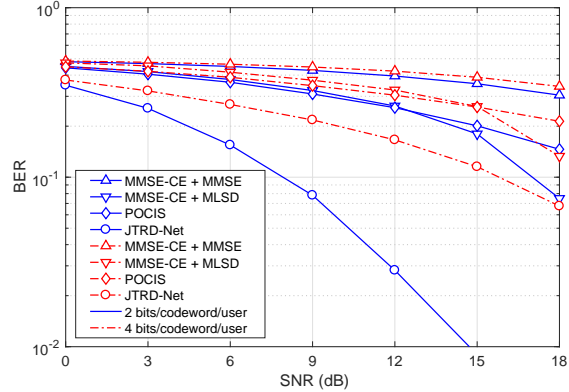}
	\caption{Average BER performance comparison of the proposed JTRD-Net approach versus other MIMO detection algorithms with different transmission rates under 8-by-8 Kronecker MIMO channels.}
	\label{figure11}
\end{figure}

\figref{figure10} illustrates the \gls{ber} performance of the JTRD-Net approach with different data rates under 4-by-8 Kronecker MIMO channels. The network is trained at \gls{snr} of 9 dB in 2 bits/codeword/user case and 12 dB in 4 bits/codeword/user case. For 2 bits/codeword/user, the gap between the JTRD-Net approach and the \gls{mmsece} with \gls{mlsd} algorithm is around 5.2 dB at \gls{ber} of $10^{-3}$. The POCIS detector achieves nearly the same performance as the \gls{mmsece} with \gls{mmse} algorithm with the performance difference around 1.2 dB. For 4 bits/codeword/user, it is shown that JTRD-Net fails to achieve full diversity gain. The gap between JTRD-Net and the \gls{mmsece} with \gls{mlsd} algorithm is approximately 4 dB at \gls{ber} of $10^{-2}$, and getting close with the \gls{snr} increasing. Meanwhile, JTRD-Net is still able to largely outperform other baseline schemes. 

\figref{figure11} illustrates the \gls{ber} performance of the JTRD-Net approach with different data rates under 8-by-8 Kronecker MIMO channels. The network is trained at \gls{snr} of 9 dB in 2 bits/codeword/user case and 15 dB in 4 bits/codeword/user case. For 2 bits/codeword/user, the POCIS detector achieves nearly the same performance as the \gls{mmsece} with \gls{mlsd} algorithm throughout the whole \gls{snr} range. The proposed JTRD-Net approach still largely outperforms all baseline schemes with at least 9 dB performance improvement. Similar phenomenons have been observed for 4 bits/codeword/user case, the performance improvement from the JTRD-Net approach to the pilot-based schemes is approximately 4 dB at high \gls{snr} regime.

\begin{figure}[!t]
	\centering
	\includegraphics[width=3.3in]{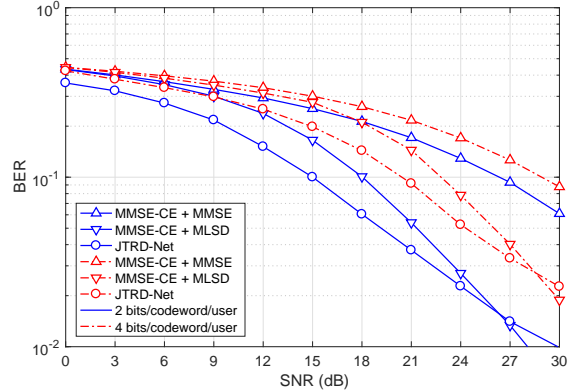}
	\caption{Average BER performance comparison of the proposed JTRD-Net approach versus other MIMO detection algorithms with different transmission rates under 4-by-8 3GPP 3D MIMO channels.}
	\label{figure12}
\end{figure}
\subsubsection{3GPP MIMO Channel}
This section aims to investigate the \gls{ber} performance of the proposed JTRD-Net approach under realistic MIMO channel models, e.g., 3GPP 3D MIMO channels. To achieve the best performance, we slightly modify the network architecture by increasing the size of the hidden-layers to 1500, 1000, and 500. Besides, we consider only the pilot-based schemes for performance comparison, since the POCIS detector cannot work in this case.

\figref{figure12} illustrates the \gls{ber} performance of the JTRD-Net approach with different data rates under 4-by-8 3GPP 3D MIMO channels. The network is trained at \gls{snr} of 21 dB in 2 bits/codeword/user case and 27 dB in 4 bits/codeword/user case. It is shown that all detection algorithms have their performance largely degraded under realistic channels, since the practical channels are very ill-conditioned and seriously correlated. For 2 bits/codeword/user, the JTRD-Net approach outperforms the conventional \gls{mmsece} with \gls{mlsd} algorithm when the \gls{snr} is lower than 27 dB. For 4 bits/codeword/user case, similar phenomenons have been observed and the proposed JTRD-Net approach outperforms the pilot-based solutions when \gls{snr} is smaller than 28.5 dB. 

\begin{figure}[!t]
	\centering
	\includegraphics[width=3.3in]{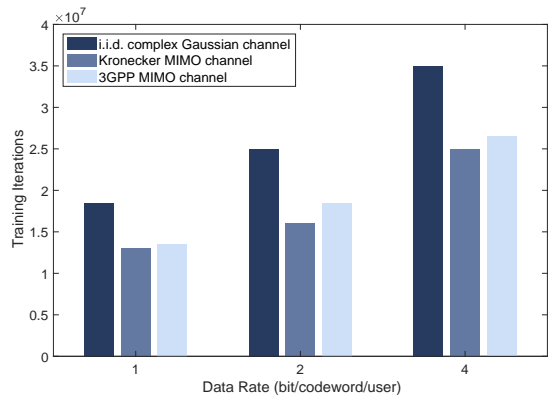}
	\caption{{\color{sblue}Number of training iterations required for various data rates under different channel models in 4-by-8 MU-SIMO system with 4 layers in DNN receiver.}}
	\label{figure13}
\end{figure}


\subsubsection{Analysis of Training Complexity}	
{\color{sblue}This section aims to investigate the training complexity of the proposed JTRD-Net under different channel models. The complexity is evaluated by comparing the required training iterations, which is obtained by continuously training JTRD-Net until the loss function converges, i.e., does not change significantly during a certain number of iterations.

\figref{figure13} illustrates the number of training iterations required for various data rates under different channel models in 4-by-8 MU-SIMO system with 4 layers in DNN receiver. Here we fix the network size and number of layers to provide a fair complexity comparison among different channel models. Note that the number of training iterations might vary largely for different training attempts. Thus, the above result is obtained by averaging 10 independent training attempts. It is shown that, for the same training set-up, i.i.d. complex Gaussian channel requires the highest training iterations, which is approximately 1.4 times than both Kronecker MIMO channel and 3GPP MIMO channel in 1 bit/codeword/user case. Similar results have been observed for 2 bit/codeword/user case and 4 bit/codeword/user case. The number of training iterations for i.i.d. complex Gaussian channel is approximately 1.6 and 1.3 times than Kronecker channel and 3GPP channel for 2 bit/codeword/user case, respectively. For 4 bit/codeword/user case, the number is approximately 1.4 times higher for both channel models. The reason that correlated channel model requires less training iterations is potential because the correlations among channel coefficients benefits the transmitter-side joint codebook design. Besides, high data rate requires more training iterations is because the size of the codebook is correspondingly larger which brings more difficulties in neural network training procedure.
}

\section{Conclusion}
In this paper, we have developed a novel end-to-end learning approach for uplink \gls{musimo} joint transmitter and non-coherent receiver design, namely JTRD-Net. The network is easy and fast to train because only feed-forward neural networks are employed. After training, the entire network can work efficiently in a non-coherent manner without requiring any channel knowledge {\color{sblue} in communication procedure}. Besides, we have developed a novel weight initialization method for JTRD-Net, which aims to mitigate the training imbalance among different UTs. Simulation results have demonstrated that the proposed JTRD-Net approach took significant advantages in terms of reliability and complexity over hand-engineered detection schemes. More interestingly, we revealed that channel correlation can benefit the deep learning-based joint transmitter and receiver design. Compared with the i.i.d. complex Gaussian channels, JTRD-Net achieved better performance and required lower computational complexity under spatially-correlated channels.

\ifCLASSOPTIONcaptionsoff
\newpage
\fi

\balance
\bibliographystyle{IEEEtran}
\bibliography{./ref}

\end{document}